\documentclass{article}
\usepackage{spconf,amsmath,graphicx}
\usepackage{booktabs}
\usepackage{enumitem}
\setlist{nosep, leftmargin=14pt}
\usepackage{hyperref}
\usepackage{mwe}

\title{Destructive augmentations for prostate lesion segmentation on heterogeneous biparametric MRI data}

\name{\begin{tabular}{@{}c@{}} Thibault Heintz$^{1}$, Rachit Saluja$^{2,1}$, Seungbin Park$^{1}$, Daniel Margolis$^{1}$, Timothy McClure$^{1}$, \\ Mert R. Sabuncu$^{1,2}$, Heejong Kim$^{1*}$ \thanks{$^{*}$Corresponding author: hek4004@med.cornell.edu}
\end{tabular}}

\address{$^{1}$Weill-Cornell Medicine, $^{2}$Cornell Tech}

\begin{document}

\maketitle

\begin{abstract}
Training prostate lesion segmentation models on biparametric MRI (bpMRI) is limited by small datasets with heterogeneous imaging acquisition protocols and annotations. In this work, we train a segmentation model on a combined cohort of $1,579$ lesion-containing training cases from five datasets, representing, to our knowledge, the largest collection of prostate lesion annotations used for this task. 

Additionally, we investigate two complementary improvements to the widely used nnU-Net framework.
First, we introduce a random wavelet augmentation and show that random sym4 (Symlet) wavelets outperform random convolution augmentation in terms of average Dice score, with only a modest reduction in F1 score due to lower precision. Second, we evaluate instance-level losses and show that Connected-Component-based Dice (CC-Dice)-CE loss complements BackSplit \cite{saluja2026backsplit}, yielding better performance than BackSplit alone.

Together, these improvements achieve the best overall performance, reaching a Dice score of 0.614 on the public Prostate158 test set and establishing a new state-of-the-art on this benchmark.

We publicly release our training code\footnote{\url{github.com/tppm/prostate\_lesion\_segmentation}} and model weights\footnote{\url{huggingface.co/heintz/prostate\_lesion\_seg}}.

\end{abstract}

\section{Introduction}
\label{sec:intro}

Biparametric MRI (bpMRI) is increasingly used in the management of suspected clinically significant prostate cancer (csPCa). Prostate lesions often appear small and have low contrast with the surrounding tissue, as the signal of malignant tissue overlaps with that of benign tissue. For radiology reporting and MRI-ultrasound fusion biopsy guidance, radiologists contour these lesions manually, a time-consuming process that shows only moderate inter-rater agreement \cite{adams2022prostate158}. Moreover, bpMRI protocols vary across hospitals. As a result, publicly available datasets are heterogeneous in both acquisition and annotation, complicating the development of robust deep learning models.

Previous studies highlight the difficulty of prostate lesion segmentation on MRI. The authors of the Prostate158 dataset \cite{adams2022prostate158} report baseline Dice scores of 0.450 and 0.400 against two expert readers. Wu et al.~\cite{wu2025lesion} outperform the baseline against both readers and report a Dice of 0.513 using arbitrated labels that select the preferred annotation from the two readers. Alzate-Grisales et al.~\cite{alzate2023sam} achieve a Dice of 0.467 using a SAM-UNETR model on a combined Prostate158 and PI-CAI cohort. Pooch et al.~\cite{pooch2026semi} train an nnU-Net model on a large cohort including PROMIS \cite{wang_2025_15683922}, Prostate158, and PI-CAI, but report only detection metrics. Direct comparison among methods is challenging as evaluation datasets, code, and model weights are often unavailable. To our knowledge, no prior study has trained a segmentation model using all currently available public datasets, including UCLA Prostate-MRI-US-Biopsy (PMUB) \cite{natarajan2020prostate}, PROMIS \cite{wang_2025_15683922}, PI-CAI including annotations by Pooch et al. \cite{pooch2026semi}, and Prostate158.

Data augmentation is a common strategy for improving robustness to variations in image appearance. Random convolutions \cite{scholz2025random} perturb image intensities and have been proposed to improve generalization by encouraging shape-based representations, enabling out-of-domain segmentation. 
Recent work by Molinier et al.~\cite{molinier2026one} similarly improves out-of-domain segmentation performance on a spine segmentation task, without a substantial decrease in in-domain performance through the careful compilation of an extensive list of augmentations, including random convolutions. 
GIN-IPA also leverages random convolutions and pairs them with a second augmentation aimed at reducing spurious correlations between items in images \cite{ouyang2022causality}. These works focus on improving generalization to out-of-domain data, such as CT to MRI generalization or generalization to different MRI sequences. However, imaging data of the same modality can still differ substantially in acquisition, for example in the choice of high b-value for diffusion-weighted imaging (DWI) in prostate MRI, the field strength and the scanner manufacturer. 

This work investigates prostate lesion segmentation on bpMRI in multi-dataset training, where cohorts differ in both acquisition and annotation. Given the acquisition differences, we suggest that random convolutions, used for generalization to out-of-domain settings, can be beneficial even in this in-domain task. Additionally, as lesion extent is ambiguous and reader-dependent \cite{adams2022prostate158}, exact alignment between the image and the ground truth may be less critical. Therefore, we propose the random wavelet augmentation, which perturbs both this image-label alignment and image intensities.

\begin{table*}[t]
\centering
\caption{An additive ablation table showing the baseline nnU-Net ResEnc-M and the cumulative effects of our contributions. The best Dice and F1 performance per dataset are bolded.}
\label{tab:additive}
\setlength{\tabcolsep}{3pt}
\resizebox{\textwidth}{!}{%
\begin{tabular}{lcccccccccccc}
\toprule
 & \multicolumn{2}{c}{All} & \multicolumn{2}{c}{Internal} & \multicolumn{2}{c}{PROMIS} & \multicolumn{2}{c}{PI-CAI} & \multicolumn{2}{c}{Prostate158} & \multicolumn{2}{c}{PMUB} \\
\cmidrule(lr){2-3} \cmidrule(lr){4-5} \cmidrule(lr){6-7} \cmidrule(lr){8-9} \cmidrule(lr){10-11} \cmidrule(lr){12-13}
 & Dice & Detection (R/P/F1) & Dice & Detection (R/P/F1) & Dice & Detection (R/P/F1) & Dice & Detection (R/P/F1) & Dice & Detection (R/P/F1) & Dice & Detection (R/P/F1) \\
\midrule
nnU-Net ResEnc-M & .391 & .624/.454/.526 & .325 & .519/.392/.447 & .341 & .560/.398/.465 & .497 & .821/.446/.578 & .609 & .870/.800/.833 & .383 & .651/.552/.597 \\
+ BackSplit gland & .402 & .594/.597/\textbf{.596} & .336 & .481/.517/\textbf{.499} & .351 & .520/.523/.522 & \textbf{.533} & .811/.706/\textbf{.755} & .603 & .826/.864/.844 & .380 & .628/.639/\textbf{.633} \\
+ CC & .408 & .635/.537/.582 & .339 & .524/.439/.478 & .364 & .593/.511/.549 & .530 & .842/.645/.731 & \textbf{.614} & .913/.808/\textbf{.857} & .388 & .640/.567/.601 \\
+ random sym4 wavelet & \textbf{.422} & .682/.492/.571 & \textbf{.346} & .535/.388/.450 & \textbf{.397} & .687/.510/\textbf{.585} & .532 & .863/.599/.707 & \textbf{.614} & .913/.808/\textbf{.857} & \textbf{.405} & .703/.492/.579 \\
\bottomrule
\end{tabular}}
\end{table*}

Dice-CE loss is a suboptimal objective for small lesion segmentation, as it is computed over all lesion voxels at once, allowing large lesions to dominate. Instance-level losses address this by considering each lesion individually, e.g., Blob loss \cite{kofler2023blob}, Connected-component-Dice-CE (CC-Dice-CE) \cite{bouteille2026learning} and Bidirectional CC (BiCC) loss \cite{bouteille2026bicc}. Complementarily, BackSplit \cite{saluja2026backsplit} splits the background into different anatomical classes, providing additional training signal that has been shown to improve segmentation of small lesions against large organ backgrounds. Combining instance-level losses with BackSplit remains unexplored. 

Our work's main contributions are: 

\begin{enumerate}
    \item Using $1,579$ lesion-containing cases, we train a segmentation model on \textbf{the largest compiled cohort used for prostate lesion segmentation} to date. Our final model achieves state-of-the-art performance on the test set of Prostate158~\cite{adams2022prostate158} with a Dice score of 0.614 against the Reader 1 annotations, compared to their 0.450 baseline.
    \item Inspired by random convolutions \cite{scholz2025random} and Random Filter Loss (RaFL) \cite{anonymous2026rafl}, we adapt \textbf{random wavelets} as an augmentation, where the MRI sequences are filtered with a randomly constructed wavelet filter. We show that these destructive augmentations can improve performance when training on heterogeneous labels and acquisition data.
    \item We evaluate \textbf{instance-level losses for prostate lesion segmentation} and show that CC-Dice-CE complements BackSplit~\cite{saluja2026backsplit}, yielding improved performance over either approach alone.
\end{enumerate}

\section{Methods}
\label{sec:methods}

We build our pipeline on nnU-Net \cite{isensee2021nnu}, a self-configuring segmentation framework for medical imaging, and use the recommended ResEnc-M preset \cite{isensee2024nnu} as our baseline. We additionally adopt BackSplit \cite{saluja2026backsplit}, adding the prostate mask to the label map, with lesion labels taking precedence. The prostate class serves as an auxiliary training target and is not used at inference. To study augmentation strategies, we compare random wavelets and random convolutions against this nnU-Net-BackSplit baseline (Sec.~\ref{sec:random_wav}). We then investigate whether instance-level losses can further improve performance with and without BackSplit (Sec.~\ref{sec:instance_loss}).

\subsection{Random wavelet augmentation}
\label{sec:random_wav}

\begin{figure}
    \centering
\includegraphics[width=\columnwidth]{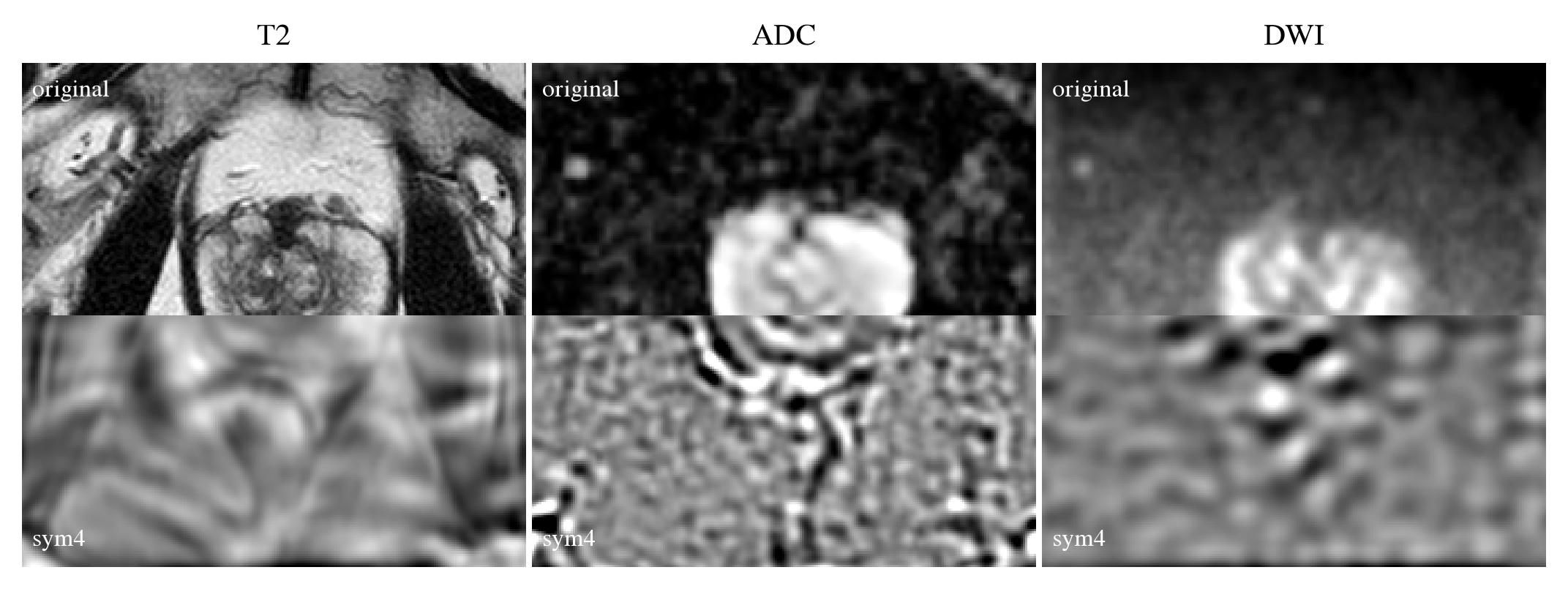}
    \caption{The sym4 filter applied to a sample of the PI-CAI dataset. Top: original sequence. Bottom: sym4 filter applied.}
    \label{fig:filter}
\end{figure}

The random wavelet augmentation replaces each sequence with the response of a randomly composed wavelet filter, with a new filter drawn per sequence. 
Each sequence is decomposed at four levels $j$ with dilations of 1, 2, 4 and 8, along the three axes $a$ using separable 1D filters with circular padding. At each level, we combine the three single-axis detail bands $\mathcal{D}^{(a)}_{j}$ along a random unit direction $u_j \sim \mathcal{U}(S^2)$, discarding the mixed-axis detail bands and low-pass residual. The level responses are then summed with random weights $v_j \sim \mathcal{N}(0,1)$, $F = \sum_{j=1}^{4}v_j\sum_{a}u_{j,a}\mathcal{D}_{j}^{(a)}$. The result is min-max rescaled to the intensity range of the original sequence. We evaluate four wavelet types: db1 (Haar, 2 taps), db2 (4 taps), db4 (8 taps, smooth but strongly asymmetric) and Symlet sym4, a near-symmetric counterpart of db4. Figure~\ref{fig:filter} shows an example of a sym4 augmentation, illustrating that the filter not only distorts textures but also displaces edges.

We compare these to the default and DA5 augmentations included in nnU-Net. The default pipeline consists of standard medical imaging augmentations, including mirroring, rotation, brightness and contrast changes, Gaussian noise and blurring. DA5 increases the strength and/or probability of the transformations and adds transformations such as blank rectangles or sharpening. 

On top of the nnU-Net default augmentations, we add either random wavelets or random convolutions with a probability of 0.5. For random convolutions, we test two configurations: 3 layers (l=3) of $3\times3\times3$ (k=3) convolutions and 5 layers (l=5) of $5\times5\times5$ (k=5). 

\subsection{Instance-level losses}
\label{sec:instance_loss}
To better account for lesion-level imbalance, we investigate augmenting the Dice and Cross-Entropy loss (Dice-CE) with instance-level losses. As instance-level losses do not benefit large components in the BackSplit class, we apply the Dice-CE loss to all classes, but limit the instance-level losses to the lesions only, as follows: 

\begin{equation}
\mathcal{L} = \mathcal{L}_{\mathrm{CE}}(\hat{Y}, Y) + \mathcal{L}_{\mathrm{SD}}(\hat{Y}, Y) + \mathcal{L}_{\mathrm{inst}}(\hat{y}_{\ell}, y_{\ell}),
\label{eq:loss}
\end{equation}

where $Y$ is the ground-truth label map for all classes and $\hat{Y}$ is the network's softmax output, $y_{\ell}$ is the binary reference lesion mask and $\hat{y}_{\ell}$ is the predicted lesion probability, i.e. the lesion channel of $\hat{Y}$, with the prostate and remaining background channel merged into a single class. $\mathcal{L}_{\mathrm{CE}}$ and $\mathcal{L}_{\mathrm{SD}}$ are nnU-Net's cross-entropy and soft Dice losses. $\mathcal{L}_{\mathrm{inst}}$ is one of the three instance-level losses described below. We use nnU-Net's deep supervision implementation and weights for every loss term.

We implement three instance-level losses: Blob loss \cite{kofler2023blob} with Dice-CE, CC-Dice-CE \cite{bouteille2026learning} and BiCC-Dice-CE \cite{bouteille2026bicc}. Blob loss iterates over all ground-truth lesions, masks out the other ground-truth lesions and calculates the loss between that lesion and the predicted mask. The results are averaged over all lesions. Alternatively, CC-Dice-CE creates Voronoi cells for each ground-truth label and calculates the Dice-CE loss within each cell. Bidirectional CC-Dice-CE (BiCC-Dice-CE) additionally divides the predicted mask into Voronoi cells and matches them to the ground-truth, reducing false positives.

\begin{table}[t]
\centering
\small
\setlength{\tabcolsep}{4pt}
\caption{Train, validation and testing data used.}
\label{tab:data_summary}
\begin{tabular}{lccc|c}
\toprule
 Dataset & Train/Val/Test & Field & High $b$ & Lesion vol \\
         &           & (T)   & (s/mm$^2$) & Median (mL) \\
\midrule
Internal & 428 / 52 / 117  & 3 & 1400 & 0.182 \\
 PI-CAI \cite{saha2024artificial} & 296 / 43 / 86 & 1.5/3 & $\geq$1000  & 1.055\\
 PROMIS \cite{wang_2025_15683922} & 317 / 33 / 87 & 1.5 & 1400  & 1.121 \\
 P158 \cite{adams2022prostate158} & 70 / 12 / 19 & 3 & 1000 & 0.441\\
 PMUB \cite{natarajan2020prostate} & 468 / 67 / 133 & 3 & 1400 & 0.347 \\
 \midrule
 Total & 1579 / 207 / 442 & & \\
\bottomrule
\end{tabular}
\end{table}

\begin{table*}[t]
\centering
\caption{The effect of all tested augmentations, starting from a baseline nnU-Net ResEnc-M configuration with a BackSplit Prostate class. Best Dice and F1 per dataset are in bold.}
\label{tab:filter_abl}
\setlength{\tabcolsep}{3pt}
\resizebox{\textwidth}{!}{%
\begin{tabular}{lcccccccccccc}
\toprule
 & \multicolumn{2}{c}{All} & \multicolumn{2}{c}{Internal} & \multicolumn{2}{c}{PROMIS} & \multicolumn{2}{c}{PI-CAI} & \multicolumn{2}{c}{Prostate158} & \multicolumn{2}{c}{PMUB} \\
\cmidrule(lr){2-3} \cmidrule(lr){4-5} \cmidrule(lr){6-7} \cmidrule(lr){8-9} \cmidrule(lr){10-11} \cmidrule(lr){12-13}
Augmentation family & Dice & Detection (R/P/F1) & Dice & Detection (R/P/F1) & Dice & Detection (R/P/F1) & Dice & Detection (R/P/F1) & Dice & Detection (R/P/F1) & Dice & Detection (R/P/F1) \\
\midrule
no random filter (nnU-Net default DA) & .402 & .594/.597/\textbf{.596} & .336 & .481/.517/\textbf{.499} & .351 & .520/.523/.522 & .533 & .811/.706/.755 & .603 & .826/.864/.844 & .380 & .628/.639/\textbf{.633} \\
DA5 & .413 & .624/.533/.575 & .339 & .508/.431/.467 & .375 & .587/.471/.522 & .536 & .842/.684/.755 & .603 & .870/.800/.833 & .395 & .628/.584/.605 \\
random conv k = 3, l = 3 & .407 & .622/.494/.551 & .329 & .492/.392/.436 & .384 & .613/.511/.558 & .526 & .832/.590/.690 & .597 & .826/.760/.792 & .385 & .628/.498/.555 \\
random conv k = 5, l = 5 & .415 & .658/.521/.581 & .340 & .514/.438/.473 & .385 & .653/.513/\textbf{.575} & .532 & .863/.636/.732 & .618 & .913/.778/.840 & \textbf{.397} & .669/.511/.579 \\
random wavelet, Haar (db1) & .411 & .621/.534/.574 & .339 & .470/.424/.446 & .383 & .640/.519/.573 & .539 & .842/.667/.744 & .624 & .870/.870/.870 & .378 & .610/.541/.574 \\
random wavelet, db2 & .415 & .646/.531/.583 & .354 & .535/.446/.486 & .380 & .633/.503/.560 & .545 & .863/.678/.759 & \textbf{.632} & .913/.840/\textbf{.875} & .378 & .622/.525/.569 \\
random wavelet, db4 & .417 & .645/.523/.577 & .345 & .508/.437/.470 & .383 & .613/.503/.553 & \textbf{.547} & .874/.680/\textbf{.765} & .619 & .913/.778/.840 & .390 & .657/.504/.571 \\
random wavelet, sym4 & \textbf{.418} & .648/.536/.587 & \textbf{.356} & .519/.442/.478 & \textbf{.386} & .627/.516/.566 & .542 & .853/.659/.743 & .615 & .913/.808/.857 & .385 & .657/.543/.595 \\

\bottomrule
\end{tabular}}
\end{table*}

\section{Experimental setup}

\subsection{Data}
\label{datasets}
We use five prostate bpMRI datasets: a dataset derived from PROMIS \cite{wang_2025_15683922}, PI-CAI \cite{saha2024artificial}, Prostate-MRI-US-Biopsy (PMUB) \cite{natarajan2020prostate}, Prostate158 \cite{adams2022prostate158} and an internal cohort, described in Table~\ref{tab:data_summary}. For Prostate158, the official test split is used. For the other datasets, we define the split ourselves and make them available as a reproducible evaluation. All datasets include three sequences: T2W, ADC and high-b-value DWI. ADC and DWI sequences are resampled to the T2W grid. For PI-CAI, we use the AI-generated prostate masks \cite{saha2023artificial}, the lesion segmentations included in the dataset and additional annotations by Pooch et al.~\cite{pooch2026semi}. For Prostate158, we use the annotations from Reader 1 \cite{adams2022prostate158}. This dataset only annotates lesions of PI-RADS $\geq4$. For our internal data we use all lesions annotated by clinicians during interpretation of the case, which generally includes only lesions with PI-RADS~$\geq3$. In PMUB, all annotated lesions correspond to lesions targeted for biopsy, while in PI-CAI, lesions with ISUP grade $\ge 2$ were annotated. As illustrated in Table~\ref{tab:data_summary}, lesion sizes vary substantially between datasets, which can be attributed to different labeling protocols. Cases without any annotated lesions are excluded.

\subsection{Training and evaluation}
We follow the nnU-Net training recipe, using 1000 epochs of 250 steps at batch size 2. All configurations are trained on the full training set described in Table \ref{tab:data_summary} for a single fold. nnU-Net's automatic configuration yields a patch size of $224\times224\times32$ voxels at a spacing of $0.5\times0.5\times1.5$mm. 

We evaluate all models with the final training checkpoint on the same held-out test sets (see Table \ref{tab:data_summary}). We report the Dice score averaged across test cases and precision, recall and F1 score calculated across all individual lesions. A predicted lesion is counted as a true positive if it overlaps a ground-truth lesion with a Dice score above $0.1$.

\section{Results}
\label{sec:results}

\textbf{Combined Improvements and Overall Performance:} 
We first evaluate the cumulative effect of our proposed modifications on top of the nnU-Net baseline.
Table~\ref{tab:additive} shows the additive effect of consecutively adding BackSplit and our two technical contributions. BackSplit \cite{saluja2026backsplit} improves Dice performance by $+.011$ and the F1 score by $+0.070$ over plain nnU-Net. CC-Dice-CE further improves this result by $+.006$ Dice, and the sym4 wavelet by another $+.014$. Both additions increase recall, but lower precision, reducing the F1 score from .596 for the BackSplit-only model to .571 for our final model. For our task, we prioritize recall over precision, as false negatives have a higher impact than false positives. Compared to the nnU-Net baseline, our final model still sees an improvement in Dice of $+0.031$ and F1 of $+0.045$.   
On the Prostate158 dataset, our final model achieves a Dice of 0.614 against the labels of Reader 1, compared to the baseline of 0.450 provided by Adams et al. \cite{adams2022prostate158} and 0.506 by Wu et al. \cite{wu2025lesion}, setting a new state-of-the-art on this dataset.

\begin{table}[h]
\centering
\caption{Comparing the effect on the Dice metric of adding an instance-level loss to the Dice-CE baseline, both with and without BackSplit. The best Dice per dataset is bolded.}
\label{tab:backsplit_loss}
\footnotesize
\setlength{\tabcolsep}{4pt}
\begin{tabular}{l|c|ccccc}
\toprule
Loss & All & Internal & PROMIS & PI-CAI & P-158 & PMUB \\
\midrule
\multicolumn{7}{l}{\textit{lesion vs background}} \\
Dice-CE & .391 & .325 & .341 & .497 & .609 & .383 \\
+ blob & .392 & .319 & .336 & .503 & .615 & .390 \\
+ CC & .398 & .328 & .354 & .500 & .615 & \textbf{.391} \\
+ BiCC & .388 & .319 & .326 & .515 & .591 & .379 \\
\midrule[0.1pt]
\multicolumn{7}{l}{\textit{BackSplit gland}} \\
Dice-CE & .402 & .336 & .351 & \textbf{.533} & .603 & .380 \\
+ blob & .401 & .331 & .351 & .524 & .611 & .385 \\
+ CC & \textbf{.408} & \textbf{.339} & \textbf{.364} & .530 & .614 & .388 \\
+ BiCC & .397 & .334 & .345 & .514 & \textbf{.624} & .379 \\
\bottomrule
\end{tabular}
\end{table}

\noindent\textbf{Effect of Destructive Augmentations:} 
Table~\ref{tab:filter_abl} compares the different augmentation strategies applied to our nnU-Net baseline optimized with BackSplit. These results show a trend that lesion segmentation performance improves as the transformations become more aggressive. While DA5 augmentations provide a substantial gain over the standard nnU-Net pipeline ($+0.011$ Dice), even larger improvements are achieved with random convolutions ($+0.013$ Dice for k=5, l=5) and long-range wavelet perturbations. Among the wavelet-based methods, Dice performance increases progressively from Haar to db2, db4, and sym4, with sym4 yielding the largest overall improvement ($+0.016$ Dice). The benefits of augmentation are most pronounced for PROMIS, the only cohort acquired entirely at 1.5T, suggesting improved robustness to underrepresented lower field strength images in our training data. 

In contrast to the improvements in Dice, the F1 score decreases with the more aggressive DA5, random convolution and wavelet augmentations. This reduction is generally driven by lower precision despite higher recall. 

For sym4 wavelets, however, the decrease is small compared to the F1 improvement obtained from BackSplit. 

These findings suggest that even highly destructive augmentations can remain beneficial in the presence of noisy labels and heterogeneous datasets. One possible explanation is that such augmentations act as a regularizer, reducing overfitting to specific image characteristics, similar to dropout in neural networks.

\noindent\textbf{Effect of Instance-Level Losses:} We evaluate three instance-level losses on top of Dice-CE, both with and without BackSplit. 
Table~\ref{tab:backsplit_loss} shows that CC-Dice-CE improves Dice performance over the Dice-CE baseline and achieves the best overall performance when combined with BackSplit. In contrast,  Blob-loss and BiCC do not provide consistent improvements across datasets.

\section{Conclusion}
\label{sec:conclusion}

In this work, we compile a large prostate lesion segmentation cohort spanning five heterogeneous datasets with substantial differences in acquisition protocols, lesion characteristics, and annotation practices. Despite this challenging setting, our final model achieves state-of-the-art performance on the Prostate158 test set, reaching a Dice score of 0.614. We find that destructive wavelet-based augmentations improve robustness to dataset heterogeneity, while CC-Dice-CE provides additional gains beyond standard Dice-CE optimization and complements BackSplit. Further investigations could study the effect of perturbing other wavelet sub-bands, different combinations of sub-bands and different wavelet types. Additionally, evaluations on different tasks and in a true out-of-domain setting could further establish the value of the random wavelet augmentations.

\section{Compliance with ethical standards}
\label{sec:ethics}

For the use of our internal data, we obtained approval of Weill-Cornell Medicine and New-York Presbyterian hospital review board. All other data are publicly accessible.

\section{Acknowledgments}
\label{sec:coi}

This work is supported by the National Institute of Health grants K25CA283145, R01AG053949, R01AG064027, and
R01AG070988. Thibault Heintz is supported by the Belgian American Educational Foundation. Daniel Margolis reports clinical advisory to Stratagen Bio/Ravada Health, holding stock options. Mert R. Sabuncu is an advisor to Cleerly Inc. and Medara Health and holds equity interests in both.
We use the AI-tool Claude for AI assisted programming and grammar checking.

\bibliographystyle{IEEEbib}
\bibliography{refs}

@inproceedings{saluja2026backsplit,
  title={Backsplit: The importance of sub-dividing the background in biomedical lesion segmentation},
  author={Saluja, Rachit and Cihangir, Asli and Deng, Ruining and Paetzold, Johannes C and Liu, Fengbei and others},
  booktitle={Proceedings of the IEEE/CVF Conference on Computer Vision and Pattern Recognition},
  pages={8492--8502},
  year={2026}
}

@inproceedings{bouteille2026learning,
  title={Learning to look closer: A new instance-wise loss for small cerebral lesion segmentation},
  author={Bouteille, Luc and Jaus, Alexander and Kleesiek, Jens and Stiefelhagen, Rainer and Heine, Lukas},
  booktitle={2026 IEEE 23rd International Symposium on Biomedical Imaging (ISBI)},
  pages={1--5},
  year={2026},
  organization={IEEE}
}

@article{bouteille2026bicc,
  title={BiCC: Bidirectional Connected-Component Loss for Instance-Aware Segmentation},
  author={Bouteille, Luc and Jonske, Frederic and Kleesiek, Jens and Jaus, Alexander},
  journal={arXiv preprint arXiv:2609.30223},
  year={2026}
}

@article{scholz2025random,
  title={Random convolutions for domain generalization of deep learning--based medical image segmentation models},
  author={Scholz, Daniel and Erdur, Ayhan Can and Peeken, Jan C and Varma, Aswathi and Graf, Robert and others},
  journal={Radiology: Artificial Intelligence},
  volume={8},
  number={1},
  pages={e240502},
  year={2025},
  publisher={Radiological Society of North America}
}

@inproceedings{kofler2023blob,
  title={Blob loss: Instance imbalance aware loss functions for semantic segmentation},
  author={Kofler, Florian and Shit, Suprosanna and Ezhov, Ivan and Fidon, Lucas and Horvath, Izabela and others},
  booktitle={International Conference on Information Processing in Medical Imaging},
  pages={755--767},
  year={2023},
  organization={Springer}
}

@article{natarajan2020prostate,
  title={Prostate MRI and ultrasound with pathology and coordinates of tracked biopsy (prostate-MRI-US-biopsy)},
  author={Natarajan, Shyam and Priester, Alan and Margolis, Daniel and Huang, Jiaoti and Marks, Leonard},
  journal={Cancer Imaging Arch},
  volume={10},
  number={7937},
  year={2020}
}

@misc{wang_2025_15683922,
  author       = {Wang, Yipei and
                  Gayo, Iani and
                  Thorley, Natasha and
                  Ng, Alexander and
                  Yan, Wen and others},
  title        = {A derived PROMIS data set},
  month        = jun,
  year         = 2025,
  publisher    = {Zenodo},
  doi          = {10.5281/zenodo.15683922},
  url          = {https://doi.org/10.5281/zenodo.15683922},
}

@article{isensee2021nnu,
  title={nnU-Net: a self-configuring method for deep learning-based biomedical image segmentation},
  author={Isensee, Fabian and Jaeger, Paul F and Kohl, Simon AA and Petersen, Jens and Maier-Hein, Klaus H},
  journal={Nature methods},
  volume={18},
  number={2},
  pages={203--211},
  year={2021},
  publisher={Nature Publishing Group US, New York}
}

@inproceedings{saha2023artificial,
  title={Artificial intelligence and radiologists at prostate cancer detection in mri—the pi-cai challenge},
  author={Saha, Anindo and Bosma, Joeran and Twilt, Jasper and van Ginneken, Bram and Yakar, Derya and others},
  booktitle={Medical Imaging with Deep Learning, short paper track},
  year={2023}
}

@article{adams2022prostate158,
  title={Prostate158-An expert-annotated 3T MRI dataset and algorithm for prostate cancer detection},
  author={Adams, Lisa C and Makowski, Marcus R and Engel, G{\"u}nther and Rattunde, Maximilian and Busch, Felix and others},
  journal={Computers in biology and medicine},
  volume={148},
  pages={105817},
  year={2022},
  publisher={Elsevier}
}

@article{saha2024artificial,
  title={Artificial intelligence and radiologists in prostate cancer detection on MRI (PI-CAI): an international, paired, non-inferiority, confirmatory study},
  author={Saha, Anindo and Bosma, Joeran S and Twilt, Jasper J and Van Ginneken, Bram and Bjartell, Anders and others},
  journal={The Lancet Oncology},
  volume={25},
  number={7},
  pages={879--887},
  year={2024},
  publisher={Elsevier}
}

@article{pooch2026semi,
  title={Semi-supervised learning in prostate MRI tumor detection approaches fully supervised performance on external validation},
  author={Pooch, Eduardo HP and Agrotis, Georgios and Cai, Lishan and Emberton, Mark and Shah, Taimur T and others},
  journal={European Radiology},
  volume={36},
  number={6},
  pages={5011--5021},
  year={2026},
  publisher={Springer}
}

@article{isensee2024nnu,
  title={nnU-Net revisited: a call for rigorous validation in 3D medical image segmentation (2024)},
  author={Isensee, Fabian and Wald, Tassilo and Ulrich, Constantin and Baumgartner, Michael and Roy, Saikat and others},
  journal={arXiv preprint arXiv:2404.09556},
  year={2024}
}

@inproceedings{molinier2026one,
  title={One Sequence to Segment Them All: Efficient Data Augmentation for CT and MRI Cross-Domain 3D Spine Segmentation},
  author={Molinier, Nathan and M{\"o}ller, Hendrik and Dagonneau, Thomas and Curto-Vilalta, Anna and Graf, Robert and others},
  booktitle={International Conference on Medical Image Computing and Computer-Assisted Intervention},
  pages={391--401},
  year={2026},
  organization={Springer}
}

@article{alzate2023sam,
  title={Sam-unetr: Clinically significant prostate cancer segmentation using transfer learning from large model},
  author={Alzate-Grisales, Jesus Alejandro and Mora-Rubio, Alejandro and Garc{\'\i}a-Garc{\'\i}a, Francisco and Tabares-Soto, Reinel and De La Iglesia-Vay{\'a}, Maria},
  journal={IEEE Access},
  volume={11},
  pages={118217--118228},
  year={2023},
  publisher={IEEE}
}

@article{wu2025lesion,
  title={Lesion-guided selective multi-modal integration for prostate cancer segmentation and PI-RADS grading in MP-MRI},
  author={Wu, Menglin and Li, Fan and Tao, Yuhui and Zhang, Yuhan and Wang, Shanshan and others},
  journal={Medical Physics},
  volume={52},
  number={10},
  pages={e70019},
  year={2025},
  publisher={Wiley Online Library}
}

@inproceedings{
anonymous2026rafl,
title={Ra{FL}: An Empirical Study of Random Filter Losses for Image Reconstruction},
author={Anonymous},
booktitle={Submitted to The Fifteenth International Conference on Learning Representations},
year={2026},
url={https://openreview.net/forum?id=KgsXW9XYqI},
note={under review}
}

@article{ouyang2022causality,
  title={Causality-inspired single-source domain generalization for medical image segmentation},
  author={Ouyang, Cheng and Chen, Chen and Li, Surui and Li, Zeju and Qin, Chen and others},
  journal={IEEE Transactions on Medical Imaging},
  volume={42},
  number={4},
  pages={1095--1106},
  year={2022},
  publisher={IEEE}
}

\end{document}